\newcommand{\ycut}{{\ensuremath{y_\mathrm{cut}}}}      
\newcommand{\pt}{{\ensuremath{p_\mathrm{t}}}}          
\newcommand{\JETSET}{{\sc jetset}}
\newcommand{\HERWIG}{{\sc herwig}}
\documentclass{ws-p8-50x6-00}

\begin{document}
\renewcommand{\thefootnote}{\fnsymbol{footnote}}    

\title{Measurement of the Scaling Property of Factorial Moments  
in Hadronic Z Decay}

\author{Gang~Chen$^ \dag$\footnotemark[1], Yuan~Hu$ ^\ddag$, LianShou~Liu$^ \dag$,
 W.~Kittel$^ \ddag$, W.J.~Metzger$^ \ddag$
}           

\address{$^ \dag$ Institute of particle physics, Huazhong Normal University, 
Wuhan, China\\
$^ \ddag$ High Energy Physics Institute Nijmegen, The Netherlands\\
E-mail: chengang@iopp.ccnu.edu.cn}


\maketitle

\abstracts{
Both three- and one-dimensional studies of local multiplicity fluctuations
in hadronic Z decay are performed using data of the L3 experiment at LEP. 
The normalized factorial moments in three dimensions exhibit power-law
scaling, indicating that the fluctuations are isotropic, which 
correspends to a self-similar fractal.
A detailed study of the corresponding one-dimensional moments confirms this conclusion.
However, two-jet subsamples have anisotropic fluctuations, 
correspending to a self-affine fractal.
These features are, at least qualitatively, reproduced by the Monte Carlo models \JETSET\ and \HERWIG.}

\footnotetext[1]{Permanent address: Jingzhou teacher's college, Hubei 434104, China.}

\section{Introduction} 
Dynamical fluctuations can be investigated using the normalized factorial moments (NFM),
\begin{equation}  
  F_q(M)=\frac {1}{M}\sum\limits_{m=1}^{M}\frac{< n_m(n_m-1) \cdots (n_m-q+1)> }{< n_m >^q}
\label{eq:spa}
\end{equation}
where $M$ is the number of bins in which momentum space is partitioned and
      $n_m$  is the multiplicity in the $m$th bin.
If  power-law scaling (intermittency),
\begin{equation}  
   F_q(M) \propto M^{\phi_q}\quad,
\label{eq:3d}
\end{equation}
is observed then the corresponding hadronic system is fractal, as is expected from a branching process.
%
%
Since particle production occurs in three-dimensional momentum space, we can distinguish two cases:
If the  scaling  is observed when the space is partitioned
equally (unequally) in each direction then the fractal is self-similar (self-affine).
For two dimensions the degree of anisotropy is quantified by the so-called Hurst exponent,
$H_{ab} = {\ln M_a / \ln M_b}$, where $M_a$ and $M_b$ are the number of bins in the two dimensions. 
Projected onto one dimension, $a$,
the second-order NFM  saturates and is given by\cite{OCHSproj}
\begin{equation}   
\label{eq:1d}
  F_2^{(a)}(M_a) = A_a - B_a M_a^{-\gamma_a} \quad.
\end{equation}

\twocolumn
\noindent
$H_{ab}$ is related\cite{Liu} to the exponents $\gamma$ through
$H_{ab} = (1+\gamma_b)/(1+\gamma_a)$.  
In three dimensions, the dynamical fluctuations are isotropic if $H_{ab}=1$  for all pairs of
axes, $ab$, or, equivalently if the $\gamma_a$ for all three axes, $a$, are equal.

Consequently, we measure the NFM's in three dimensions
to examine whether there is power-law scaling 
and to measure the intermittency indices, $\phi _q$. 
We also measure the one-dimensional  $F_2$
to get the exponents $\gamma_a$ for the three directions.

To  investigate possible  dependence of the fluctuations on perturbative or  non-perturbative scales,
we  also examine the $F_2$ for  2-jet events as a function  of 
\ycut.

\section{Method of analysis} 

Since we are studying  dynamical fluctuations, the coordinate system should not
depend on the dynamics of $\mathrm{q} \bar\mathrm{q} \rightarrow\mathrm{hadrons}$. 
So we choose
 a cylindrical coordinate system with $z$ axis along the $\mathrm{q}\bar\mathrm{q}$
 direction and we partition the phase space in bins of $y$, \pt, $\varphi$.
The $y$ and \pt\ directions are defined by the electro-weak process.
But, we need an origin for $\varphi$. Clearly,  we should not use an axis, like major,
which depends on the QCD dynamics.  One possibility\cite{diffsh} is
to rotate the coordinate system around the
thrust axis and to define a new $x$ axis as lying in the $z$-beam plane.
However, it is easier simply to choose the origin  at random.

\begin{figure}
 \begin{center}
  \epsfig{figure=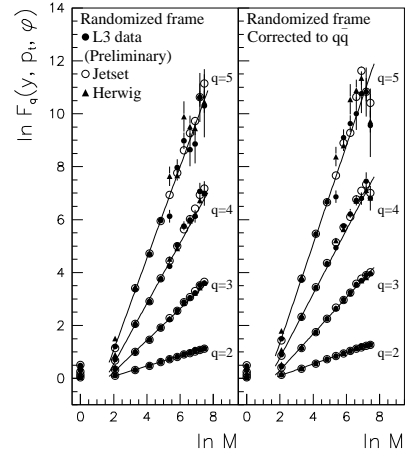,width=.445\textwidth,bbllx=32,bblly=22,bburx=342,bbury=375}
 \end{center}
 \caption{
 The 3D NFM as a function of the number of bins, $M$,
 compared with  results of \JETSET\ and \HERWIG.
 \label{fig:3d}
 }
\end{figure}

\begin{figure}
 \begin{center}
  \epsfig{figure=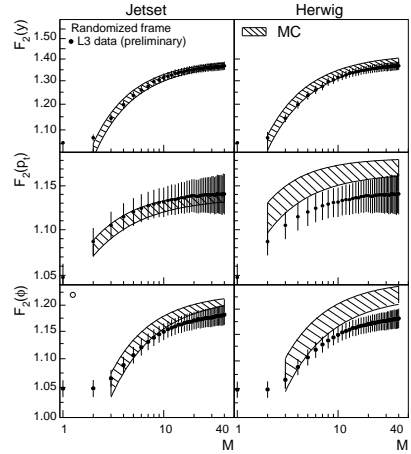,width=.447\textwidth,bbllx=1,bblly=17,bburx=453,bbury=529}
 \end{center}
 \caption{
 The $F_2$ in one dimension as function of number of bins,  $M$, in the
 randomized frame,
 compared to \JETSET\ and \HERWIG.
 \label{fig:1d} 
}
\end{figure}

\onecolumn
Experimentally, we estimate the
$\mathrm{q} \bar\mathrm{q}$ direction by the thrust axis.
We also examine the effect of correcting this direction to the
$\mathrm{q} \bar\mathrm{q}$ direction using a correction factor determined from \JETSET.

The data used in the analysis were collected by the L3 detector   
in 1994 at 
the Z pole.
Hadronic events are selected and cuts are applied to
obtain well measured tracks and calorimetric clusters.
A total of about one million events satisfy the selection criteria.
 
The NFM's are calculated from the raw data and corrected for detector effects by
a  factor determined from
generator level MC and  detector level MC.
Systematic errors are estimated from different selection cuts, different
methods of event selection and different MC for detector correction.

\begin{table} 
\caption{The fit parameters of the three-dimensional NFM}
\label{tab:3d}
\begin{center}
 \begin{tabular}{|c||c|c||c|c|}\hline
\ & \multicolumn{2}{|c||}{Randomized}
  & \multicolumn{2}{|c|}{Corrected to $\mathrm{q}\bar\mathrm{q}$ } \\ \hline
 order  &$\phi_q $&$ \chi^2$/dof  & $\phi_q $&$ \chi^2$/dof \\ \hline
 2 &$0.194\pm0.003\pm0.003$ & 8/9 & $0.221\pm0.003\pm0.003$ & 8/9 \\
 3 &$0.598\pm0.011\pm0.014$ &11/9 & $0.685\pm0.011\pm0.012$&11/9 \\
 4 &$1.082\pm0.013\pm0.018$ & 8/9 & $1.206\pm0.022\pm0.026$& 7/9 \\
 5 &$1.731\pm0.024\pm0.025$ & 6/9 & $1.858\pm0.028\pm0.035$& 7/9 \\  \hline
\end{tabular}
\end{center}
\end{table}

\vskip-0.3cm

\section{Results of \boldmath{$F_q$} for the full sample} 
The three-dimensional $F_q$ for an isotropic partitioning of phase space,
are shown in Fig.\,\ref{fig:3d} together with the results of fits of Eq.\,\ref{eq:3d},
omitting the first point to eliminate the influence of momentum 
conservation.  The results of the fit are also given in Table\,{\ref{tab:3d}.
The fits are good, as is expected if
dynamical fluctuations are isotropic.
Correcting from the thrust to the
$\mathrm{q} \bar\mathrm{q}$ direction  systematically increases the values of the $\phi_q$.
Both \JETSET\ and \HERWIG\ are seen to agree well with the data.

The  one-dimensional $F_2$ are plotted in Fig.\,\ref{fig:1d}
and the results of  fits  of  Eq.\,\ref{eq:1d} are shown in Table\,\ref{tab:1d}. 
The values of $\gamma$ for the three axes are equal, confirming that the fluctuations are isotropic.
Correcting 
to the
$\mathrm{q} \bar\mathrm{q}$ direction  decreases slightly the $\gamma$ values.
Again, \JETSET\ and \HERWIG\  agree  with the data.

\section{Results from 2-jet subsamples} 

We use the Durham algorithm to define 2-jet events for various values of the
jet resolution parameter, \ycut.
We  fit the resulting one-dimensional $F_2$ using 
\twocolumn[
Eq.\,\ref{eq:1d}
and plot the values of $\gamma$ vs.\   
$k_\mathrm{t} = \sqrt{s\ycut}$ in Fig.\,\ref{fig:2jet}.
While $\gamma$  increases
with $k_\mathrm{t}$ (or \ycut) for each of the variables,  the dependence on \ycut\ is very different for
the different variables,  $y$, \pt, and $\varphi$.
 \JETSET\ and \HERWIG\ behave qualitatively similarly to the data.

\section{Conclusions} 
In three dimensions,  the factorial moments exhibit power-law
scaling when momentum space is partitioned
isotropically. 
Fits to the three-dimensional $F_q$ vs.\ $M$ determine the intermittency indices $\phi_q$.
In one dimension,  fits to $F_2(y)$, $F_2(p_{t})$, $F_2(\varphi)$ confirm the isotropy.
Both \JETSET\ and \HERWIG\ agree remarkably well with
both the three-dimensional $F_q$ and the one-dimensional $F_2$.
For 2-jet events, $F_2(y)$, $F_2(p_{t})$, $F_2(\varphi)$ depend very
differently on the jet resolution parameter. 
Similar behavior is seen in \JETSET\ and \HERWIG.
\vspace{9mm}
]


\begin{table} 
\caption{The fit parameters of the 1-D NFM.}
\label{tab:1d}
\begin{center}
 {\tabcolsep5pt
\begin{tabular}{|c||c|c|}\hline
    \multicolumn{3}{|c|}{Randomized}  \\ \hline
             &  $\gamma$                & $\chi^2\!/\mathrm{dof}$ \\ \hline
       $y$   &  $0.992\pm0.014\pm0.028$ & $36/36$ \\
         \pt &  $0.986\pm0.022\pm0.032$ & $29/37$ \\
   $\varphi$ &  $0.993\pm0.024\pm0.033$ & $25/35$ \\ \hline\hline
    \multicolumn{3}{|c|}{Corrected to $\mathrm{q}\bar\mathrm{q}$}  \\ \hline
             &  $\gamma$                & $\chi^2\!/\mathrm{dof}$ \\ \hline
       $y$   &  $0.966\pm0.019\pm0.018$ & $35/36$ \\
         \pt &  $0.972\pm0.042\pm0.030$ & $31/37$ \\
   $\varphi$ &  $0.967\pm0.034\pm0.024$ & $27/35$ \\ \hline
\end{tabular}
 }
\end{center}
\end{table}


\begin{figure} 
  \begin{center}
  \vspace{3mm}
  \epsfig{file=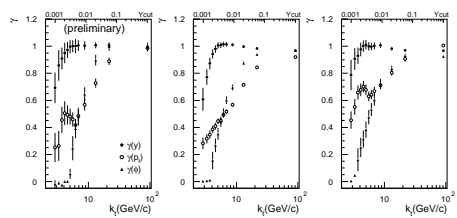,width=.5\textwidth,bbllx=34,bblly=44,bburx=629,bbury=325} 
  \end{center}
 \caption{
 The variation of $\gamma_i$ ($i$ = $y$, $p_t$, $\varphi$)
 with $k_\mathrm{t}$ (\ycut) for data (left), \JETSET\ (middle) and \HERWIG\ (right).
 \label{fig:2jet}
 }
\end{figure}

\end{document}